\documentclass[pra,twocolumn,showpacs]{revtex4-1}
\usepackage{amsmath,color,bm}
\usepackage{amsfonts,amssymb,graphicx}

\def\t{^\mathrm{T}}

\begin{document}

\title{Multipartite continuous-variable entanglement distillation using local squeezing and only one photon-subtraction operation}
\author{Song Yang$^1$, XuBo Zou$^1$, ShengLi Zhang$^1$, Bao-Sen Shi$^1$, Peter van Loock$^{23}$ and GuangCan Guo$^1$}\email{xbz@ustc.edu.cn}
\affiliation{1 Key Laboratory of Quantum Information, University of Science and
Technology of China (CAS), Hefei 230026, China.\\
2 Optical Quantum Information Theory Group, Max Planck Institute for the Science of Light, G\"unther-Scharowsky-Str.1/Bau 26, 91058 Erlangen, Germany\\
3 Institute of Theoretical Physics I, Universit\"at Erlangen-N\"urnberg, Staudstr.7/B2, 91058 Erlangen, Germany
 }
\date{\today}

\begin{abstract}
In this paper, we study entanglement distillation of multipartite continuous-variable Gaussian entangled states. Following
Opatrn\'{y} \emph{et al.}'s photon subtraction (PS) scheme,
the probability of successful distillation decreases exponentially with the number of parties $N$. However, here, we shall propose an entanglement distillation scheme whose success probability scales as a constant with $N$. Our protocol employs several local squeezers, but it requires only a single PS operation. Using the logarithmic negativity as a measure of entanglement, we find that both the success probability and the distilled entanglement can be improved at the same time. Moreover, an $N$-mode transfer theorem (transferring states from phase space to Hilbert space) is presented.
\end{abstract}

\maketitle


Entanglement, particularly multipartite entanglement state, is one of the most fundamental and puzzling aspect in quantum mechanics. However, entanglement is such a fragile resource that it may be easily degraded during its interaction with the environmental noise. To this point, entanglement distillation (always in a probabilistic way) has been proposed to increase the entanglement in the noise-disturbed entangled state\cite{Hage08,Dong}. Restricted by
the famous No-Go theorem in continuous variable (CV) entanglement distillation\cite{nogo1,nogo2,nogo3}, lots of efforts have been devoted to the non-Gaussian operations. As an example, photon subtraction(PS) operation\cite{Opatrny}, proposed by Opatrn\'{y} \emph{et al} in 2000, is principally simple and can be readily implemented with beamsplitter and photon detectors. Very recently, about 10 years after Opatrn\'{y} \emph{et al}'s pioneering work, an experiment which faithfully implements the PS-based two mode entanglement distillation has been reported \cite{NatPhotonic}. One of the challenge in this experiment is the extremely low successful probability, which is mainly due to the rather-high-transmittance beamsplitter used in PS operation---For one thing, the beamsplitter must own a relatively high transmittance to  guarantee an entanglement-enhanced distillation\cite{distill2copy}. For the other, high transmittance means low reflectiveness in beamsplitter and hence, low probability in PS operation. Assuming the beamsplitter's transmittance is $0.90$, as shown in Ref.\cite{NatPhotonic}, detectors 's efficiency $0.10$, the probability of each successful local PS operation is upper-bounded by $10^{-2}$. The successful probability of the whole entanglement distillation will be even lower, decreasing exponentially with the number $N$ of local PS operations: $10^{-2N}$. This is really a serious problem if $N$-partite continuous variable entanglement distillation is involved.

In this paper, we consider the distillation of $N$-partite CV gaussian entanglement state with only one-time photon subtraction. Generally for $N$ partite (especially, the symmetric) Gaussian state, one-time PS will modify the permutation symmetry and the distilled entanglement, measured with logarithmic negativity (log-neg), will be even worse. However, if assisted with local squeezing, we show that the entanglement can be improved\cite{cv2density}. Moreover, one-time PS can give a substantial increment in distilling probability, which keeps constants $O(10^{-2})$ for arbitrary partite number $N$.

 \begin{figure*}[htp]
  \includegraphics[width=12cm]{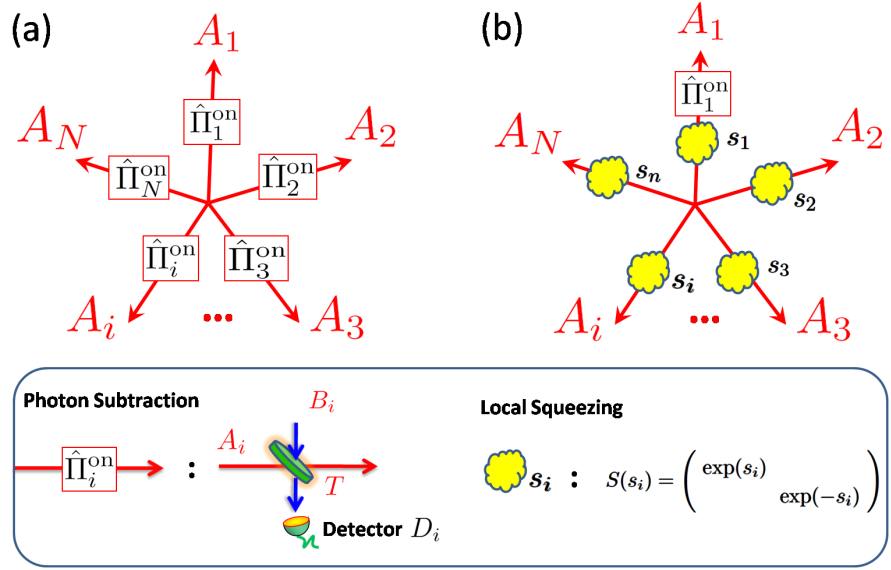}\\
  \caption{Multi-partite Entanglement distillation scheme with Photon Subtraction technique. (a)Typical PS-based $N$-partite distillation with $N$ times photon subtraction. The successful distillation is heralded by the events when all the detectors register non-vacuum result, with the probability scaling as $10^{-2N}$. (b) One-time PS and local squeezing assisted entanglement distillation. The success probability conditioned on only-one detector's detection result keep fixed in $O(10^{-2})$. }\label{schemeDis}
\end{figure*}

Our entanglement distilling scheme is briefly shown in Fig.\ref{schemeDis}, where Fig.\ref{schemeDis}(a) describes the typical $N$-time PS scheme which is a straightforward generalization of Opatrn\'{y} \emph{et al}'s scheme to $N$ partite case. Fig.\ref{schemeDis}(b) is the our
new distilling protocol with local squeezing ( described by symplectic operation $S(s_i)$ ) and only one-time PS. Throughout this paper, for convenience, we assume that the second input-mode $B_i$ of beam-splitter are vacuum modes, representing a simple and concise expression of PS operation in Phase space(See Appendix\ref{secPS}).

\emph{Preliminaries.} Our result can be conveniently derived in Phase space. Let's now introduce the basic facts and notation of CV $N$ modes state. First, it is convenient to express each mode, say $k$ mode, with the field quadrature operators $\hat{x}_k=(\hat{a}_k+\hat{a}_k^\dagger)/\sqrt{2}, \hat{p}_k=(\hat{a}_k-\hat{a}_k^\dagger)/(i\sqrt{2})$, with $\hat{a}_k , \hat{a}_k^\dagger$ being the mode annihilation and creation operators. By defining the vector of quadrature operators $\hat{X}\equiv (\hat{x}_1, \hat{p}_1,\cdots,\hat{x}_N, \hat{p}_N)$, the commutation relation can be written as $[\hat{X}_m,\hat{X}_n]=i\Omega_{mn}$, with $\Omega=\bigoplus _{k=1}^N
\left(
\begin{array}{cc}
0 & 1 \\
-1 & 0
\end{array}
\right).$
The density matrix of arbitrary $N$-mode system $\rho$ which resides in infinite-dimensional Hilbert space can be conveniently represented by the characteristic function in $2N$-dimensional real vector space, i.e., phase space
$\chi(\xi)=
\mathrm{Tr}[\rho\exp(i\hat{X}\t\xi)],\xi\in\mathbb{R}^{2N}$. Gaussian state is a special kind of quantum state whose characteristic function is Gaussian in phase space,
$
\chi (\xi) =\exp \left[ -\frac{1}{2}\xi \t%
V\xi+i\mathbf{\bar{x}}\t \xi\right],\xi\in\mathbb{R}^{2N}
$, inwhich $\mathbf{\bar{x}}$ is quadrature average $\mathbf{\bar{x}}=\mathrm{Tr}[\hat{X}\rho]$ and
$V$ denotes the covariance matrix $V_{lm}
=\frac{1}{2}\langle \hat{X}_l\hat{X}_m+\hat{X}_m\hat{X}_l\rangle-\langle \hat{X}_l\rangle\langle\hat{X}_m\rangle
$. In the following, we also use the wigner function, the Fourier transform of $\chi(\xi)$
to express the state evolution. A standard normalized wigner function is defined by
$W(\mathbf{r})=\int_{\mathbb{R}^{2N}}\frac{d^{2N}
\xi}{{(2\pi)}^{2N}}\exp\left[-i\mathbf{r}\t
\xi\right]\chi(\xi),
$ with $\vec{r}_i\in \mathbb{R}^{2}$ and $\mathbf{r}=(\vec{r}_1,\vec{r}_2,\cdots,\vec{r}_N)\in\mathbb{R}^{2N}$.

\emph{Multi-partite entanglement states.} Multipartite CV Gaussian entanglement plays a prominent role in future quantum network and quantum communication protocol\cite{jzhang,adesso}. In this paper, we are mainly interested in the distilling of a family of genuinely $N$ partite symmetric Gaussian entangled state\cite{loockInequa,loockTHEO,aokiEXP}. The covariance matrix is given by
\begin{eqnarray}
V_N=\pmb{\epsilon} \otimes |\mathcal{I}\rangle\langle \mathcal{I}|+(\pmb{\alpha}-\pmb{\epsilon})\otimes I_N,\label{symmNGenuine}
\end{eqnarray}
inwhich $|\mathcal{I}\rangle$ is the unnormalized $N$ dimensional real vector $|\mathcal{I}\rangle=(1,1,1\cdots, 1)^\mathrm{T}\in \mathbb{R}^n$, and $I_N$ is the identity matrix in $N$ dimensions. $\pmb{\alpha}=diag(a,b),\pmb{\epsilon}=diag(c,d)$ are $2\times 2$ diagonal matrix
\begin{eqnarray}
a&&=\frac{1}{2N}\left(e^{2r_1}+(N-1)e^{-2r_2}\right),c=\frac{1}{2N}(e^{2r_1}-e^{-2r_2}),\nonumber\\
b&&=\frac{1}{2N}\left(e^{-2r_1}+(N-1)e^{2r_2}\right),d=\frac{1}{2N}(e^{-2r_1}-e^{2r_2}).
\end{eqnarray}
The genuinely $N$ partite state in Eq.(\ref{symmNGenuine}) can be experimentally prepared with a particular sequence of $N-1$ phase-free beam splitters and $N$ squeezed input states\cite{loockInequa}. To be simple, we will mainly focused on the unbiased states, namely
$r_1=r_1(r_2)=\frac{1}{2}\ln[(N-1)\sinh(2r_2)]+\frac{1}{2}\ln[\sqrt{1+[(N-1)\sinh(2r_2)]^{-2}}+1]$.

\emph{Distillation with local squeezing and one-time PS.} Let's now derive the state evolution of the our one-time PS distilling protocol.
 As shown in Fig.\ref{schemeDis}(b), $N$ local squeezing symplectic transformations $S(s_i)_{i=1,\cdots, N}$ are applied before PS. This corresponds in phase space to a  transformation of covariance matrix $V_N\rightarrow V_N^S=[\oplus_{i=1}^N S(s_i)] V_N [\oplus_{i=1}^N S(s_i)\t]$. Without loss of generality, we assume that only PS operation is performed in the $A_1$ mode. Implemented with a beam-splitter (transmittance $T$), the PS operation (Fig.\ref{schemeDis}(b) inset), couples the $N$ mode Gaussain state $V_N^S$ with the vacuum mode $B_1$. The $(N+1)$ mode states now follows
 $V_{BS}=B(V_N^S\oplus\frac{1}{2}I_2)B\t$, with $B$ being the symplectic matrix
$ B=\left(
 \begin{array}{ccc}
 \sqrt{T}I_2&&-\sqrt{R}I_{2}\\
 &I_{2(N-1)}&\\
\sqrt{R}I_2 &&\sqrt{T}I_2
 \end{array}
 \right).
$
Finally, a successful distillation is heralded if the detector register non-vacuum results. According to Appendix\ref{secPS}, one can find the distilled state is a linear combination of two gaussian state
\begin{eqnarray}
\rho_{\mathrm{dis}}=\frac{\delta}{\delta-1}\rho(\Gamma_1)-\frac{1}{\delta-1}\rho(\Gamma_2),\label{distilled}
\end{eqnarray}
with $\delta=\sqrt{\det(\Gamma_2+I_2/2)}$ and $\rho(\Gamma)$ being a normalized $N$-paritie gaussian state with covariance matrix $\Gamma$. The $\Gamma_1$ and $\Gamma_2$ are defined by partitioning of matrix $V_{BS}\equiv\left(\begin{array}{cc}
\Gamma_1&M\\
M\t&\Delta
\end{array}\right),\Gamma_2=\Gamma_1-M(\Delta+I_2/2)^{-1}M\t$, where $\Gamma_1,M,\Delta$ are $2N\times 2N, 2N\times 2, 2\times 2$ matrice respectively.
The success probability of distilling follows $P_{\mathrm{succ}}=(\delta-1)/\delta$.
With the transfer theorem form phase space to Hilbert space (See Appendix \ref{secPhase2Hilbert}), one can easily compare the entanglement before and after distillation.

\begin{figure}
  \includegraphics[width=8.5cm]{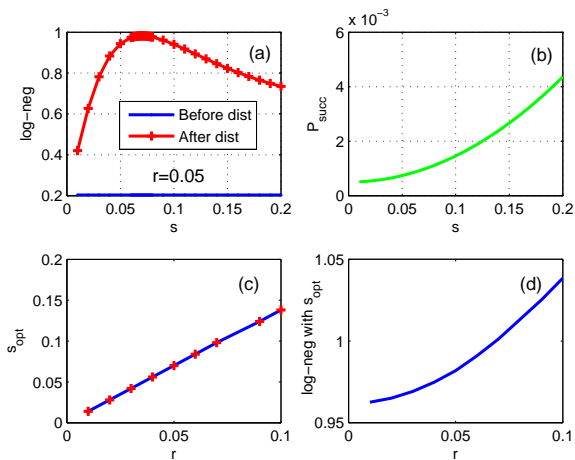}\\
  \caption{Entanglement before and after distillation for $N=3$ partite genuine Gaussian entanglement state (Eq.(\ref{symmNGenuine})) with $r_2=r=0.05,r_1=r_1(r_2)=0.099$ and $s_1=s_2=s_3=s,T=0.9$.(a)Entanglement distillation of unbiased entangled state. There
  exists an optimal local squeezing $s$ which optimizes the distilled entanglement. For $r=0.05$, $s_{opt}=0.07$ and corresponding maximal log-neg and success probability is $0.9818$ and $9.67\times 10^{-3}$. (b) The probability of successful distillation as a function of squeezing $s$. (c)Optimized local squeezing $s_{opt}$ as a function of initial squeezing parameter $r$. $s_{opt}$ scales linearly with the initial parameter $r$: $s_{opt}\approx 1.4 r$. (d) The corresponding
 entanglement with $s_{opt}$. In the numerical simulation, we truncate the photon number of each mode at $D=7$, namely, we consider only the contribution of $|0\rangle,|1\rangle,|2\rangle,\cdots |6\rangle$. By choosing $D=7$, the error during our state transfer can be controlled with in $5\times 10^{-5}$.
}\label{svaries} %
\end{figure}

\begin{figure}
  \includegraphics[width=8.5cm]{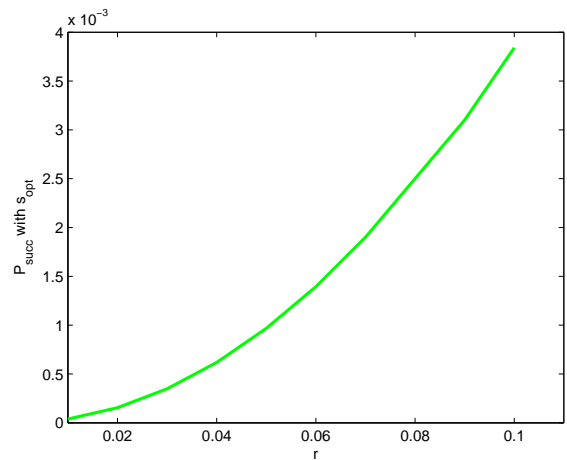}\\
  \caption{All parameters are choosen as in Fig.\ref{svaries}(d). Only here we use the optimal local squeezing $s_{opt}$ and plot the success probability.}\label{optsuccP}
\end{figure}

In Fig.\ref{svaries}, we evaluate the entanglement after and before distillation with the log-neg \cite{logneg} as the figure of merit for entanglement. For simplicity, we consider three-mode entanglement distillation as the example. For a fixed initial squeezing $r=0.05$, as shown in Fig.\ref{svaries}(a), we plot the entanglement as a function of $s_1=s_2=s_3=s$. Through our simulation, we assume the transmittance of PS beamsplitter is $T=0.90$ which is available in recent experiments\cite{NatPhotonic}. The probability of success is briefly shown in Fig.\ref{svaries}(b). It should be noted that the success probability which is about $O(10^{-3})$. This is mainly due to the rather low initial squeezing ($r=0.05$), which results extremely low photon number in each transmission mode, which certainly decrease the probability of being photon subtracted. Our method can be applied for even stronger squeezing. In Fig.(c), we increase $r$ and find the optimal squeezing $s_{opt}$ which may maximize the log-neg of output entanglement state. The numerical results support the linearity reliance of $s_{opt}$ upon the increasing $r$. Also, we plot the corresponding optimized log-neg and success probability in Fig.\ref{svaries}(d) and Fig.\ref{optsuccP}. The success probability (Fig.\ref{optsuccP}) is about $O(10^{-3})$ which is an pronounced improvement compared with the $3$-time PS strategy($O(10^{-9})$).

\emph{Discussions and Prospectives.}
We presented here a photon-subtraction based entanglement distillation for arbitrary $N-$ partite continuous variable entanglement state. As an example, now in this paper, only the three-partite symmetric Gaussian state is involved. This method is applicable for arbitrary $N-$ partite CV state. Indeed, even for $N=2$, this improvement in both log-neg and success probability also applies.
As an auxiliary result, we also derive the transfer theorem for $N-$ partite Gaussian state from Phase space to Hilbert space. We can envisage that this theorem could find more application in the entanglement evaluation tasks, such as entanglement swapping and entanglement distribution.

\emph{Acknowledgements.}
The authors acknowledge financial support from  National Fundamental Research Program, also by National Natural
Science Foundation of China (Grant No. 10674128 and 60121503) and
the Innovation Funds and \textquotedblleft Hundreds of
Talents\textquotedblright\ program of Chinese Academy of Sciences
and Doctor Foundation of Education Ministry of China (Grant No.
20060358043). SZ acknowledges support by the Max Planck Gesellschaft, Chinese Academy of
Sciences Joint Doctoral Promotion Programme (MPG-CAS-DPP). PvL acknowledges support from the Emmy Noether Program of the DFG.

\appendix
\section{Wigner function description of Photon Subtraction}\label{secPS}
In ideal cases, a perfect photon subtraction is described with the annihilation operation: $a=\sum_{n=1}^\infty \sqrt{n} |n-1\rangle\langle n|$ in Hilbert space. However, this is not an unitary operation and cannot be implemented deterministically. A convenient way is to use beamsplitter and photon detectors\cite{Opatrny}. For ease, we consider a single partite state $\rho_i$ as input, the PS operation
$\hat{\Pi}_i^{\mathrm{on}}$(see Fig.\ref{schemeDis})
can be represented with a completely-positive map from $\rho_{i}$ to normalized output state $\widetilde{\rho_{i}}=\mathcal{E}(\rho_{i})/\mathrm{Tr}[\mathcal{E}(\rho_{i})]$, with
\begin{eqnarray}
\mathcal{E}(\rho_i) &&=\mathrm{Tr}_{_B}\left[U_{_{AB}} \left(\rho_{i} \otimes |0\rangle _{_{B_i}}\langle 0|\right) U_{_{AB}}^\dagger \left(I_{_{A_i}}\otimes \hat{\Pi}_{i}^{\mathrm{on}}\right)\right]\label{psope}
 \end{eqnarray}
where
$\hat{\Pi}_{i}^{\mathrm{on}}=\sum_{n=1}^\infty|n\rangle_{_{B_i}}\langle n|$ denotes the positive operators projecting to non-vacuum subspace and $U_{_{AB}}= \exp[\arccos(\sqrt{T}) (a_{_{A_i}}^\dagger a_{_{B_i}}-a_{_{A_i}}a_{_{B_i}}^\dagger )]$ denotes the Beam-splitting operation between $A_i$ and $B_i$ modes.

In our calculation, indeed, it is convenient to use the wigner function to describe the PS process above (Eq.(\ref{psope})). In fact, the operator $\hat{\Pi}_{i}^{\mathrm{on}}=I-|0\rangle_{_{B_i}}\langle 0|$ is a difference of two operation whose wigner function are both Gaussian\cite{PRLCVCHSH}, i.e.,
\begin{eqnarray}
W(\vec{r}_i,\hat{\Pi}_{i}^{\mathrm{on}})=\frac{1}{2\pi}\left(1-2\exp[\vec{r}_i I_2 \vec{r}_i^{\mathrm{T}}]\right)
 \end{eqnarray}
In case that the input state $\rho_i$ is Gaussian, the wigner-function of distilled entanglement can be easily formulated with the linear
combination of a series of Gaussian function, each of which can be conveniently expressed with the covariance marices.

\section{Quantum State from phase-space to Hilbert Space}\label{secPhase2Hilbert}

In this section, we give the detailed techniques we use in the processing of multi-partite Gaussian quantum state from phase-space to
Hilbert Space. For a $N$-partite CV state, the
density matrix follows
\begin{eqnarray}
\rho=\int \frac{d\vec{\mu}}{\pi^N} \mathrm{Tr} \left[\rho D(\vec{\mu})\right]D(-\vec{\mu}), \vec{\mu}=(\mu_1,\mu_2,\cdots,\mu_N)\in\mathbb{C}^N\nonumber
\end{eqnarray}
with
$D(\vec{\mu})=\exp\left[\vec{\mu}\vec{a}^\dagger-\vec{\mu}^*\vec{a}\right]=\exp\left[(\vec{\mu},\vec{\mu^*})(\vec{a}^\dagger,-\vec{a})^{\mathrm{T}}\right]$ being the $N$ mode displacement operator.

The matrix entries of $\rho$ can be conveniently obtained by observing the equation\cite{cv2density}:
\begin{eqnarray}
&&\langle k_1,k_2,\cdots,k_N|D(-\vec{\mu})|m_1,m_2,\cdots, m_N\rangle\nonumber\\
=&&\frac{\prod_{i=1}^N \partial_{t_i}^{k_i}\prod_{i=1}^N \partial_{t_i^\prime}^{m_i} }{\sqrt{\prod_{i=1}^N k_i! \prod_{i=1}^N  m_i!}}\exp\left[\vec{t}\vec{t^\prime}-\vec{t}\vec{\mu}+\vec{t^\prime}\vec{\mu^*}\right]\vline_{\vec{t}=\vec{t^\prime}=0}\nonumber
\end{eqnarray}
where $\vec{t}=(t_1,t_2,\cdots,t_N),\vec{t^\prime}=(t_1^\prime,t_2^\prime,\cdots,t_N^\prime)$ is the $N$-dimensional real vector.

By noticing the fact that
\begin{eqnarray}
&&(\vec{a}^\dagger,-\vec{a})=i L_N(\hat{x}_1,\hat{p}_1,\hat{x}_2,\hat{p}_2,\cdots,\hat{x}_N,\hat{p}_N)^\mathrm{T} \\
 && L_N =\left(
\left(
\begin{array}{cc}
\frac{-i}{\sqrt{2}} &\frac{-1}{\sqrt{2}}\nonumber\\
 \frac{i}{\sqrt{2}} & \frac{-1}{\sqrt{2}}
\end{array}
\right)\otimes I_N\right)P,
P_{kl}=\begin{cases}
\delta_{2k-1,l} & k\le N\\
\delta_{2(k-N),l}   &  k > N
\end{cases}
\end{eqnarray}
one obtains that (after integrating the $\vec{\mu}$)
 \begin{eqnarray}
&&\langle k_1,k_2,\cdots,k_N|\rho|m_1,m_2,\cdots, m_N\rangle\nonumber\\
=&&\frac{\prod_{i=1}^N \partial_{t_i}^{k_i}\prod_{i=1}^N \partial_{t_i^\prime}^{m_i} F \vline_{\vec{t}=\vec{t^\prime}=0}}{\sqrt{\prod_{i=1}^N k_i! \prod_{i=1}^N  m_i!}}\label{rhoDerive}
  \end{eqnarray}
inwhich
\begin{eqnarray}
 F&&=\exp\left[\frac{1}{2}(\vec{t},\vec{t^\prime})R(\vec{t},\vec{t^\prime})^\mathrm{T}\right]/\sqrt{\det(\Gamma+I_{2N}/2)}\\
R &&=\sigma_x\otimes I_N+\left(\sigma_z\otimes I_N\right) L_N^{*}(\Gamma+I_{2N}/2)^{-1}L_N^{\dagger}\left(\sigma_z\otimes I_N\right) \nonumber \\
\end{eqnarray}
and $\sigma_x,\sigma_z$ is the Pauli matrices.
Then, one can check that the state $\rho$(Eq.(\ref{rhoDerive})) is now automatically normalized, i.e., $\mathrm{Tr}[\rho]=1$.


\begin{thebibliography}{99}
\bibitem{Hage08} B.\ Hage,
Aiko Samblowski, James Diguglielmo, Alexander Franzen,
Jarom\'{i}r Fiur\'{a}{\v{s}}ek  and Roman Schnabel,
Nat. Phys. \textbf{4}, 915 (2008).
\bibitem{Dong} R.\ Dong,
Mikael Lassen, Joel Heersink, Christoph Marquardt, Radim Filip,
Gerd Leuchs and Ulrik L Andersen,
Nat. Phys. \textbf{4}, 919 (2008).


\bibitem{nogo1} J. Eisert, S. Scheel, M. B. Plenio,
Phys. Rev. Lett, \textbf{89}, 137903 (2002).
\bibitem{nogo2} J.  Fiur\'{a}{\v{s}}ek,
Phys. Rev. Lett. \textbf{89}, 137904 (2002).
\bibitem{nogo3} G. Giedke, J. I. Cirac,
Phys. Rev. A \textbf{66}, 032316 (2002).

\bibitem{Opatrny} T. Opatrn\'{y}, G. Kurizki and D. G. Welsch,
 Phys. Rev. A \textbf{61}, 032302 (2000).
\bibitem{NatPhotonic} H. Takahashi, J. S. Neergaard-Nielsen, M. Takeuchi, M. Takeoka, K. Hayasaka, A. Furusawa, and M. Sasaki,
Nature Photonic \textbf{4}, 178 (2010).

\bibitem{distill2copy}ShengLi Zhang, Peter van Loock, Phys. Rev. A \textbf{82}, 062316 (2010).
\bibitem{cv2density} ShengLi Zhang, Peter van Loock, arXiv:1103.4500v1 .

\bibitem{jzhang} Jing Zhang, Gerardo Adesso, Changde Xie and Kunchi Peng, Phys. Rev. Lett. \textbf{103}, 070501(2009).
\bibitem{adesso} Gerardo Adesso, and Fabrizio Illuminati, Phys. Rev. A. \textbf{78}, 042310(2008).


\bibitem{loockInequa} Peter van Loock and Akira Furusawa, Phys. Rev. A \textbf{67}, 052315(2003).
\bibitem{loockTHEO} P. van Loock, and Samuel L. Braunstein,
Phys. Rev. Lett. \textbf{84}, 3482 (2000).
\bibitem{aokiEXP} Takao Aoki, Nobuyuki Takei, Hidehiro Yonezawa, Kentaro Wakui, Takuji Hiraoka, Akira Furusawa and Peter van Loock,
Phys. Rev. Lett. \textbf{91}, 080404 (2003).

\bibitem{logneg} G.\ Vidal and R.F.\ Werner, 
Phys. Rev. A \textbf{65}, 032314 (2002).

\bibitem{PRLCVCHSH} R. Garc\'{i}a-Patr\'{o}n, J. Fiur\'{a}{\v{s}}ek, N. J. Cerf, J. Wenger, R. Tualle-Brouri, and Ph. Grangier,
Phys. Rev. Lett. \textbf{93}, 130409 (2004).

\end{thebibliography}
\end{document}